# WebNC: efficient sharing of web applications

Laurent Denoue, Scott Carter, John Adcock, Gene Golovchinsky, Andreas Girgensohn
FXPAL
3400 Hillview Ave., building 4
Palo Alto, CA 94304, USA
1(650)842-4821

{denoue,carter,adcock,gene,andreas}@fxpal.com

## ABSTRACT
WebNC is a browser plugin that leverages the Document Object Model for efficiently sharing web browser windows or recording web browsing sessions to be replayed later. Unlike existing screen-sharing or screencasting tools, WebNC is specially optimized to handle scrolling within a web page. Rendered pages are captured as image tiles, and transmitted to a central server through http post. Viewers can watch the webcasts in real-time or asynchronously using a standard web browser: WebNC only relies on html and javascript to reproduce the captured web content. Along with the visual content of web pages, WebNC also captures their layout and textual content for later retrieval. The resulting webcasts require very little bandwidth, are viewable on any modern web browser including the iPhone and Android phones, and are searchable by keyword.

## Categories and Subject Descriptors
H.5.3 [INFORMATION INTERFACES AND PRESENTATION]: Group and Organization Interfaces, Web-based interaction

## 1. INTRODUCTION
Screencasting and screensharing are useful for real-time or asynchronous collaboration [1]. They were designed to capture desktop windows [2,4], not specifically web pages. But today a lot of user interaction happens inside a web browser window, so we designed WebNC (Figure 1) specifically to capture web applications.

Living inside a browser as a plugin, WebNC has access to the Document Object Model of the pages, including the type of its elements such as text, pictures, video clips, flash movies, applets, and the location of the scrollbars in these elements. By leveraging this data, WebNC can efficiently capture a rendered web page as image tiles along with metadata like viewport size, cursor position and scroll positions, and send them to a central server.

Once captured, these webcasts can then be viewed using a standard web browser without requiring any plugin. WebNC outputs its webcasts as standard HTML and Javascript, making them viewable with most modern web browsers, including smartphones such as the Apple iPhone and Google Android.

Because it uses only outbound HTTP requests for both the plugin and the viewing javascript code, WebNC avoids the need to reconfigure the network firewall. Unlike most video-codec based screen-recorders, WebNC captures webcasts at their native resolution, thus producing high-resolution and readable webcasts.

Unlike existing tools, WebNC also captures textual elements being rendered by the web browser and their location on screen, allowing users to retrieve webcasts by content.

Finally, WebNC solves a real privacy problem that emerges with existing tools: with WebNC, users can share just a single TAB in their browser, and any popup not part of that TAB will not be shared, like this private instant message or email popping up during a web-conference.

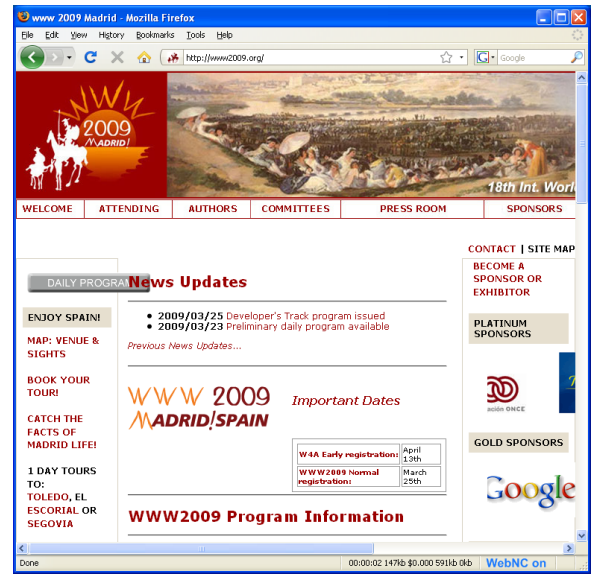

**Figure 1. On the bottom right, the WebNC icon on reads "WebNC on" and is blue when the user shares a web browser window; otherwise, it reads "WebNC off" and turns black. Session time and bandwidth is shown left of the icon.**

## 2. WebNC SYSTEM DESCRIPTION
### 2.1 User experience
To start sharing a web page, the user simply clicks on the WebNC icon on the bottom/right corner of the browser window (after having installed the WebNC extension which works on all platforms, including Windows, Mac and Linux). The icon changes color to indicate that it is actively recording. The user is shown a sessionID that can be given to others for them to view the live webcast, or kept for later viewing.

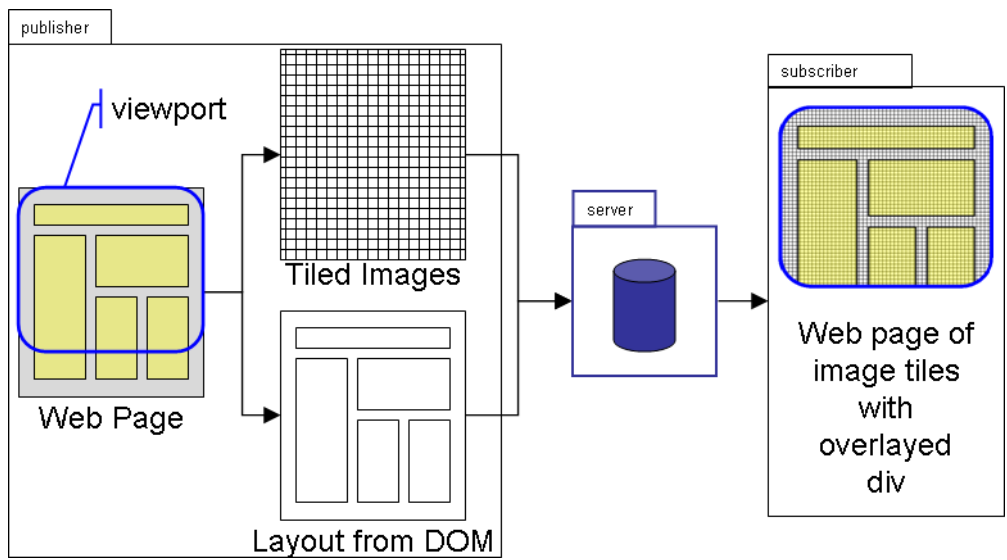

**Figure 2: WebNC system overview. The extension (publisher) captures the visual content of the web page as a set of tiled images along with its context (viewport, layout, etc..). The server then acts as a proxy, sending image tiles, layout information, and scroll events to the viewer (subscriber).**

By default, WebNC only captures the browser tab that was active when the user initiated the session. If the user switches to another tab in the browser, this new tab will not be captured. The user needs not worry about moving the window outside of the screen boundary as WebNC will still be able to capture these regions (see next section). When the user is done sharing, a simple click on the WebNC icon ends the session.

## 2.2 Capturing web pages

Living inside the browser, the WebNC plugin has access to the Document Object Model (DOM). It determines the vertical and horizontal positions of the scrollbars and uses them to name the tiles of 256x256 pixels that it captures, starting from the top/left corner of the viewable area on the web page. Tiles are captured using the drawWindow function of the canvas element in Firefox [3]. An equivalent call is available in Internet Explorer. Each tile is named after its position on the page, starting from 0,0 for the top/left tile and moving right and down until it reaches the entire scrollable area, not only the viewable area. WebNC has also access to the hidden pixels, so it can grab entire 256x256 tiles even when some of the pixel rows and columns fall outside of the viewable area.

Every once in a while, WebNC grabs the entire viewable area as a reference bitmap using drawWindow API from the canvas object [3]. It then compresses into PNG or JPG each tile until all tiles in the viewable area are compressed. Images are hashed into a unique signature comprising the MD5 of the URL, bits of the image itself, and the tile number. Tiles that have not been sent already are sent, along with the cursor position, its shape, and the size of the window, scroll position and scrollable area. This data is also sent even when no tile has changed, giving the server the latest cursor position and scrolling information.

## 2.3 Capturing text

WebNC also queries the nsIAccessibleText object of text elements to extract characters and their bounding boxes. This content is optionally sent along with the tile images and stored on the server. Users can retrieve specific parts of webcasts containing a given keyword. Privacy settings can filter out common entities such as email addresses, social security numbers, phone numbers, and street addresses.

## 2.4 Viewing webcasts: html and javascript

A standard web browser pointed to the WebNC server is enough to view a session. The page embeds javascript code that uses Ajax polling to retrieve the metadata of tile names and viewport size, scroll position, and cursor data.

The javascript creates a DIV element set to the viewport width and height, and embeds another DIV set to the size of the scrollable area inside the first DIV. It then creates as many IMG elements are required, setting their SCR to the tile values received earlier. Finally, the position of the embedded DIV is set to the scroll positions, giving users the effect of scrolling the viewport to the corresponding position. The mouse position is used to place another IMG absolutely over the first DIV element, with its SRC set to replicate the shape of the cursor.

Because the viewer code receives the data periodically from the server, WebNC uses javascript timers to smoothly transition the locations of the fake cursor as well as the scroll positions. This technique makes for a very smooth and enjoyable viewing experience.

## 3. EVALUATION and DISCUSSION

Our current system is built as an extension for Firefox, and thus runs on Windows, Mac and Linux. For Windows, we also built a native C++ plug-in to capture objects such as Flash/QuickTime/Applets for which drawing surfaces were not available to the drawWindow function on Window. We built the WebNC server using Java/Tomcat with a simple JSP pages. Data sent from the plugins are stored in memory, keyed under a given sessionid. The viewer code is in pure html and javascript and thus works on all web browsers tested so far, including Firefox,

Internet Explorer, Safari, as well as mobile web browsers based on WebKit as found in iPhone and Android G1 smartphones.

A few users tried WebNC and could easily share their browser window. Viewers were equally able to view webcasts, and were quite impressed with the smooth cursor and scrolling behaviors.

Using WinMacro [5], we recorded a 2.5 minute web browsing session involving navigation among several Wikipedia articles and the Fuji Xerox homepage. In all, 9 unique web pages were visited in the course of the session. The interactions included a fair amount of scrolling. We replayed the session in a full-screen web browser window on a 1024x768 desktop, viewed it remotely with various methods, and measured the network use by capturing the network traffic with WireShark [6]. WebNC used an average of 280 kbps, versus 130 kbps for Microsoft Remote Desktop Protocol (RDP), 521 kbps for UltraVNC with tight encoding and caching enabled and 729 kbps for Microsoft SharedView. We ran a follow-up test of RDP where we disabled the persistent bitmap cache and rebooted the client to insure that the in-memory tile cache was empty. In this second test the average bandwidth used by RDP was measured at 225 kbps.

WebNC is not yet as bandwidth efficient as RDP, though leveling the playing field by starting from an empty image cache as described above narrows the gap significantly. The current implementation of WebNC has several opportunities for optimization. First, sending back only tiles that have changed to the client could save up to 150 kbps. Second, network overhead can further be reduced by changing our polling Ajax to a server-push comet technique (*e.g.*: long polling). It's worth restating that unlike RDP and the others, WebNC does not require software beyond a standard web browser to view shared content.

## 4. CONCLUSION AND FUTURE WORK

WebNC works and has been shown to produce low bandwidth webcasts that are compliant with web standards for viewing. The ability to retrieve webcasts by content is extremely interesting, and more work is required. It would be possible to allow users to retrieve webcasts based not only on keywords captured during the webcast, but also on element types such as whether the page contained forms, video clips, images, applets, etc. WebNC needs to be optimized to cut out dynamic elements like movie clips and treat these as separate tiles, or sprites.